\documentclass{article}
\usepackage{glas}
\usepackage{graphicx}
\graphicspath{{eps/}{pics/}{figs/}}
\begin{document}

\begin{titlepage}{GLAS-PPE/97--04}{27 August 1997}

\title{The effects of radiation on Gallium Arsenide radiation detectors}

\author{R.L.~Bates\InstAnotref{glas}{\dagger}
C.~Da'Via\Instref{glas}
S.~D'Auria\Instref{udine}
V.~O'Shea\Instref{glas}
C.~Raine\Instref{glas}
K.M.~Smith\Instref{glas}}
\Instfoot{glas}{Dept. of Physics \& Astronomy, University of Glasgow, UK}
\Instfoot{udine} {Dip. di Fisica, Universit\`a di Udine, Italy}
\Anotfoot{\dagger}{Partially supported by a CASE award from the Rutherford Appleton Lab , UK}

\collaboration{On Behalf of the RD8 Collaboration}

\begin{abstract}

Semi-insulating, undoped, Liquid Encapsulated Czochralski (SI-U LEC) GaAs detectors have been irradiated with 1MeV neutrons, 24GeV/c protons, and 300MeV/c pions. The maximum fluences used were 6, 3, and 1.8~10$^{14}$ particles/cm$^{2}$ respectively. For all three types of irradiation the charge collection efficiencies (cce) of the detector are reduced due to the reduction in the electron and hole mean free paths. Pion and proton irradiations produce a greater reduction in cce than neutron irradiation, with the pions having the greatest effect.
The effect of annealing the detectors at room temperature, at 200$^{o}$C and at 450$^{o}$C with a flash lamp have been shown to reduce the leakage current and increase the cce of the irradiated detectors. The flash-lamp anneal produced the greatest increase in the cce from 26\% to 70\% by increasing the mean free path of the electrons.
Two indium-doped samples were irradiated with 24GeV/c protons and demonstrated no improvement over SI U GaAs with respect to post-irradiation cce.

\end{abstract}
\end{titlepage}

\section{Introduction}

The possible use of GaAs detectors in the future ATLAS experiment at the LHC imposes severe demands on the radiation hardness of the GaAs detectors.
Semi-insulating, undoped GaAs detectors have been shown to withstand doses up to 100Mrad of $^{60}$Co gamma rays \cite{gamma}. Their sensitivity to neutrons, protons and pions are reported in this paper.

The irradiations were performed at the ISIS facility \cite{isis} for the neutron exposure, with a spectrum strongly peaked at 1MeV, the CERN PS for 24GeV/c protons \cite{ps} and the PSI, Villigen, for 300MeV/c pions \cite{psi}. All irradiations were performed at room temperature.

\section{The detectors}

All the detectors (except for two described in section \ref{sec:indoped}) were fabricated with SI U GaAs substrates \cite{mcp}, with thicknesses varying from 115$\mu$m to 500$\mu$m. The substrates were lapped and polished at Glasgow University to obtain the desired thickness. Various thicknesses were used so that the dependence of the leakage current and charge collection efficiency on this parameter could be studied. All the detectors had 3mm diameter circular pad contacts with a 200$\mu$m wide guard ring separated from the pad by 10$\mu$m. The Schottky contact on one face of the wafer was metallized with Ti/Pd/Au and the ohmic contact on the other face was Pd/Ge annealed at 300$^{o}$C.
The leakage current characteristics before and after the irradiations were obtained with a Keithley 237 voltage source measurement unit. A $^{90}$Sr source was used to determine the response of the detectors to minimum ionising particles with an amplifier shaping time of 500ns. The charge collection efficiencies of either electrons or holes created in the first 20$\mu$m of the detector thickness were obtained with the use of an $^{241}$Am alpha particle source placed on the Schottky or ohmic contact, respectively. An amplifier with a 500ns shaping time was used. The current characteristics and charge collection efficiencies from the $^{90}$Sr source were obtained at both 20$^{o}$C and -10$^{o}$C. The -10$^{o}$C temperature was chosen as this is the proposed operating temperature of the ATLAS semiconductor tracker. 

\section{Reverse Current}

\begin{figure}
\centering
\resizebox{.7\textwidth}{!}{\includegraphics{{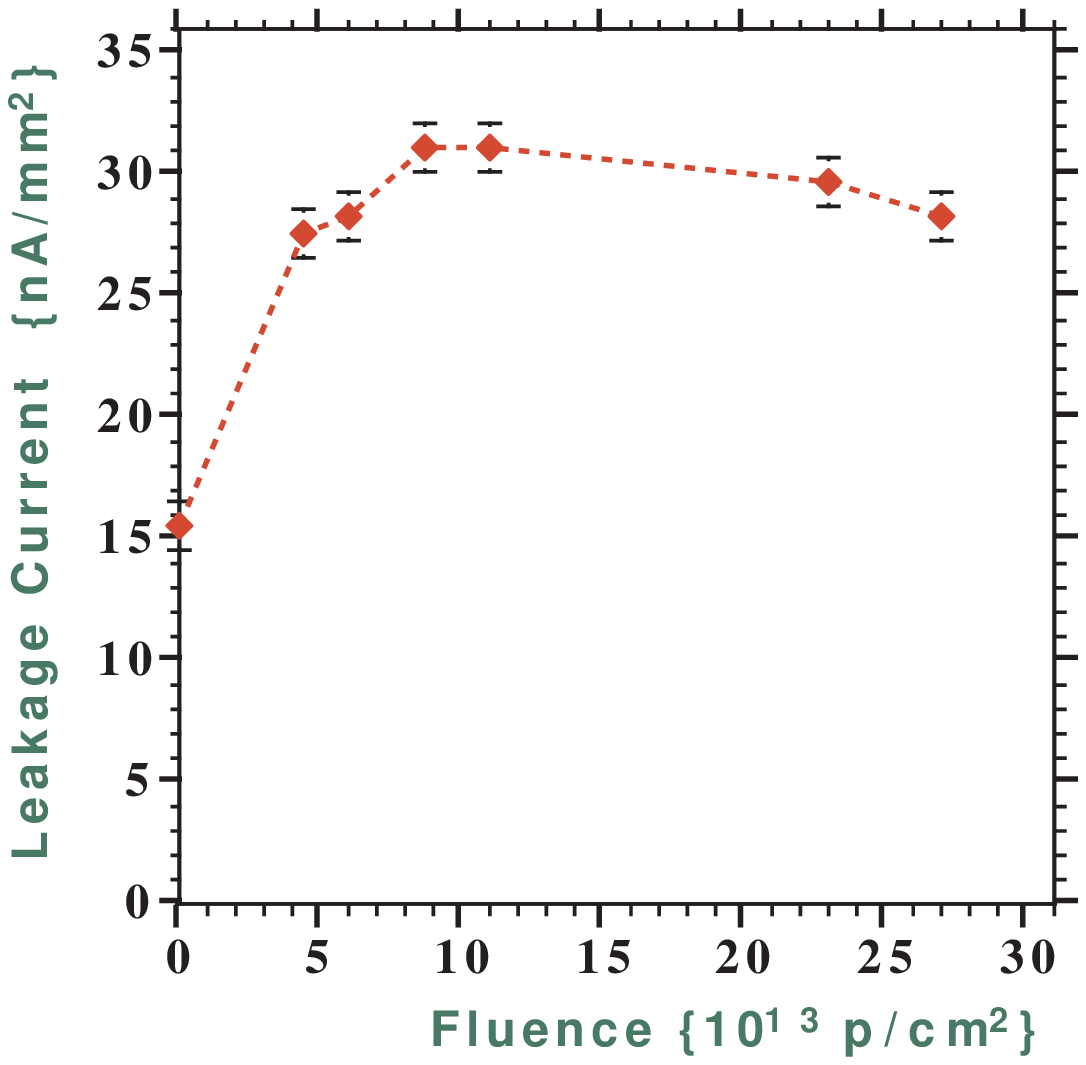}}}
\caption{The reverse current of a 160$\mu$m thick GaAs detector as a function of fluence of 24GeV/c protons, measured at 20$^{o}$C.
\label{fig:ivdose}}
\end{figure}

The effect of all three types of irradiation on the reverse current characteristics of the GaAs detectors is small.
\begin{figure}
\centering
\resizebox{.7\textwidth}{!}{\includegraphics{{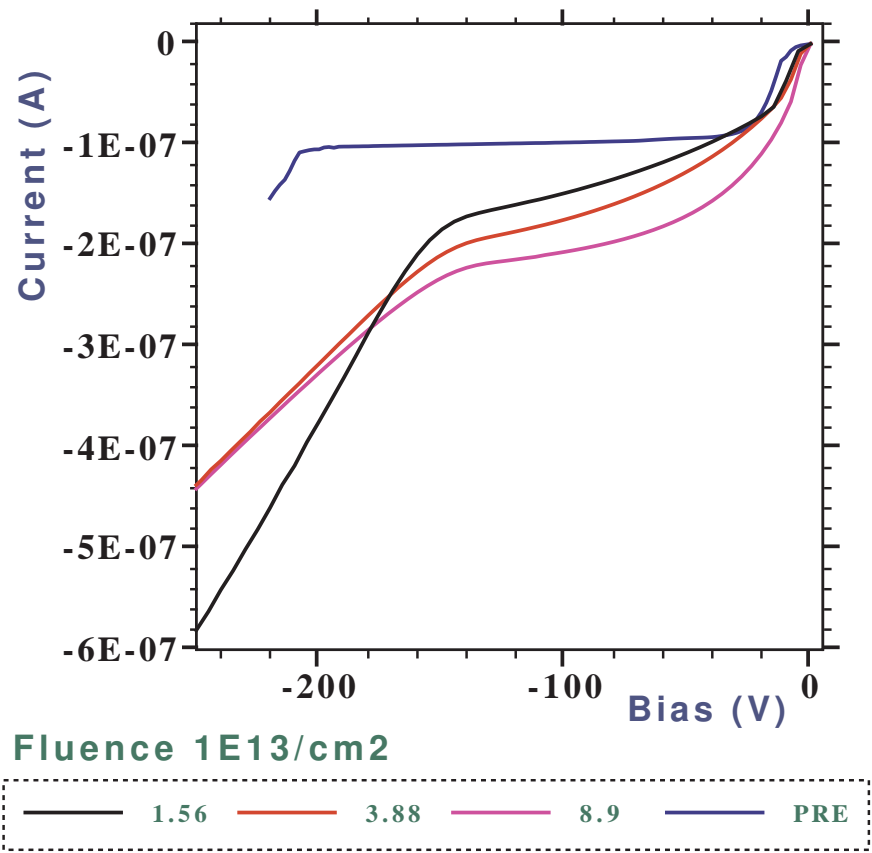}}}
\caption{The reverse current characteristics of a 210$\mu$m thick GaAs detector with increasing 24GeV/c proton fluence, measured at 20$^{o}$C
\label{fig:iv}}
\end{figure}
Figure \ref{fig:iv} shows four reverse current characteristics with increasing proton fluence for a 210$\mu$m thick detector measured at 20$^{o}$C. The plateau current after a fluence of 8.9~$\times$~10$^{13}$~p/cm$^{2}$ is only 2.2 times larger than the pre-irradiated value. Before irradiation the detector is fully depleted at a bias, V$_{fd}$, of 250V; after only 1.56~$\times$~10$^{13}$~p/cm$^{2}$ this has reduced to 160V. The value remains unchanged with higher fluence. For an applied bias greater than V$_{fd}$ the signal created by irradiating the ohmic contact with an alpha source (the hole signal) is non-zero, implying that the detector is active across the entire bulk. As the value of V$_{fd}$ is reduced with fluence the extension of the high field region in the detector changes after irradiation.
The value of the reverse current for biases greater than V$_{fd}$ reduces with increased fluence and the reverse current characteristic becomes more resistive in nature. Although the reverse current at V$_{fd}$ increases with dose up to  1~$\times$~10$^{14}$~p/cm$^{2}$, the majority of the increase occurs by 3~$\times$~10$^{13}$~p/cm$^{2}$. From 1~$\times$~10$^{14}$~p/cm$^{2}$ up to  2.4~$\times$~10$^{14}$~p/cm$^{2}$ (the maximum fluence tested with this sample) the current at this bias remains approximately constant (see figure \ref{fig:ivdose}).

\begin{figure}
\centering
\resizebox{.7\textwidth}{!}{\includegraphics{{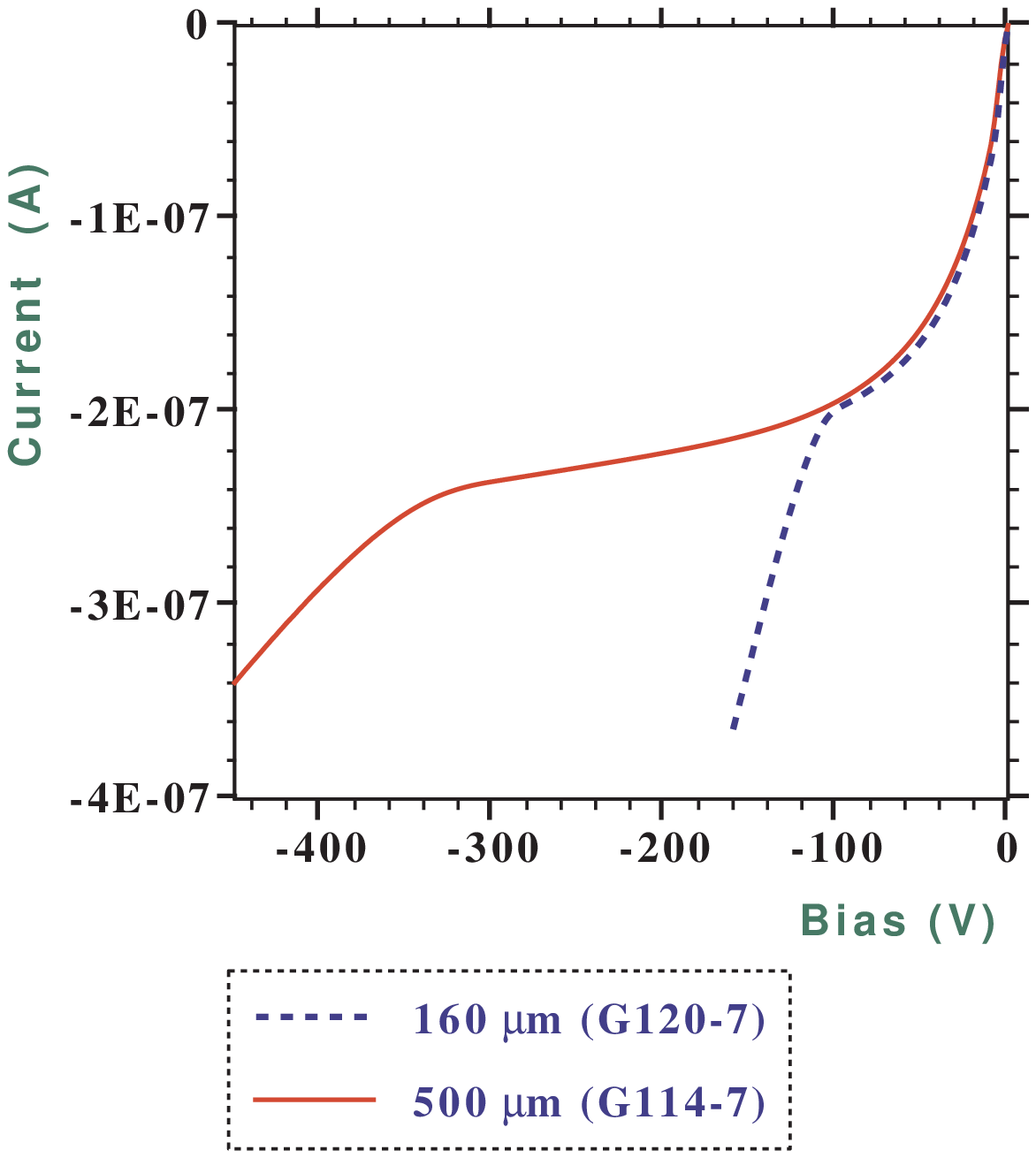}}}
\caption{The reverse current characteristics of two GaAs detectors after 6~$\times$~10$^{13}$ 24GeV/c protons/cm$^{2}$. 
\label{fig:thickiv}}
\end{figure}

Reverse current characteristics of two detectors of different thickness, 160$\mu$m and 500$\mu$m, both of which were irradiated to 6~$\times$~10$^{13}$~p/cm$^{2}$ are shown in figure \ref{fig:thickiv}. It can be seen that the value of the leakage current up to V$_{fd}$ is independent of the detector thickness. This is not the case in silicon where the leakage current, I$_{r}$, has a  dependence on fluence, $\Phi$, given by \cite{alpha-s}
\begin{equation}
\frac {\Delta I_{r}}{Vol} = \alpha \Phi 
\end{equation}
where $\alpha$ is the reverse current damage constant and Vol is the volume of the detector. No varriation in this effect on the leakage current has been observed for detectors that were biased during irradiation.

\begin{figure}
\centering
\resizebox{.7\textwidth}{!}{\includegraphics{{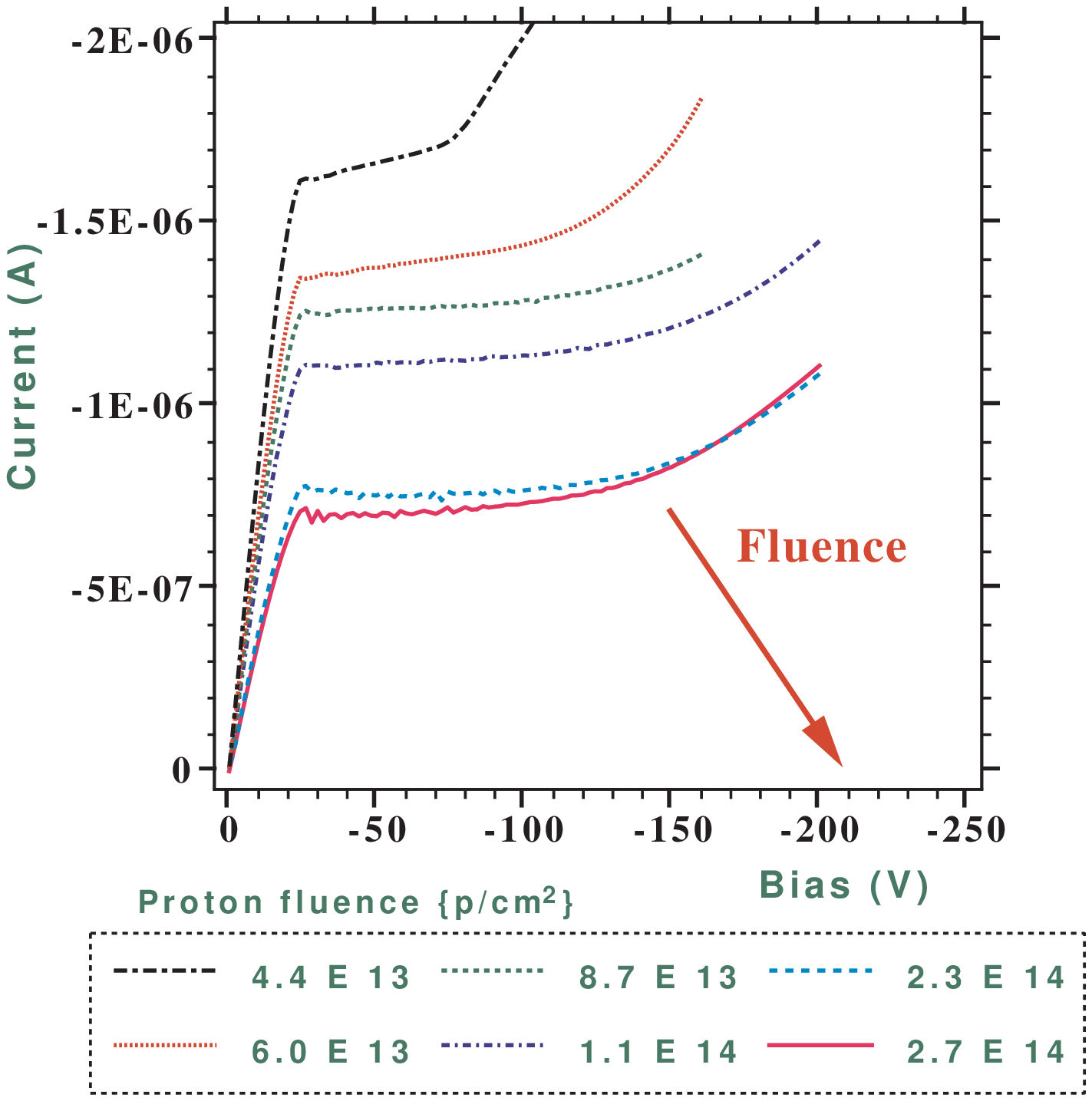}}}
\caption{The forward current characteristics of a 160$\mu$m thick GaAs detector with increasing 24GeV/c proton fluence, measured at 20$^{o}$C
\label{fig:forwardiv}}
\end{figure}

The forward current characteristics also change with increasing fluence. The effect of protons on a 160$\mu$m thick detector, shown in figure \ref{fig:forwardiv}, illustrates this. A plateau appears that extends to higher bias values with increasing fluence while the magnitude of the plateau current falls. This allows the detector to be operated under forward bias.

\section{Charge collection efficiency}

\begin{figure}
\centering
\resizebox{.7\textwidth}{!}{\includegraphics{{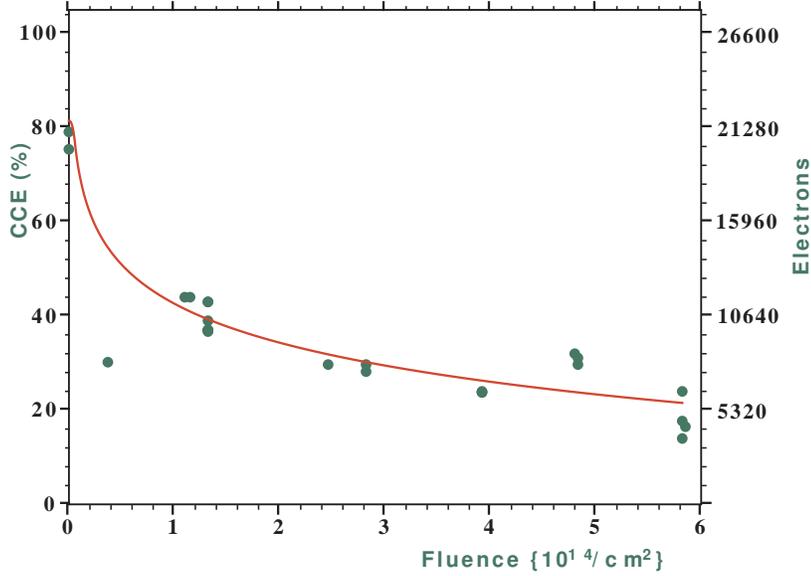}}}
\caption{The cce of 200$\mu$m thick GaAs detectors as a function of neutron fluence, measured at 200V and 20$^{o}$C. 
\label{fig:cceneutrons}}
\end{figure}

\begin{figure}
\centering
\resizebox{.7\textwidth}{!}{\includegraphics{{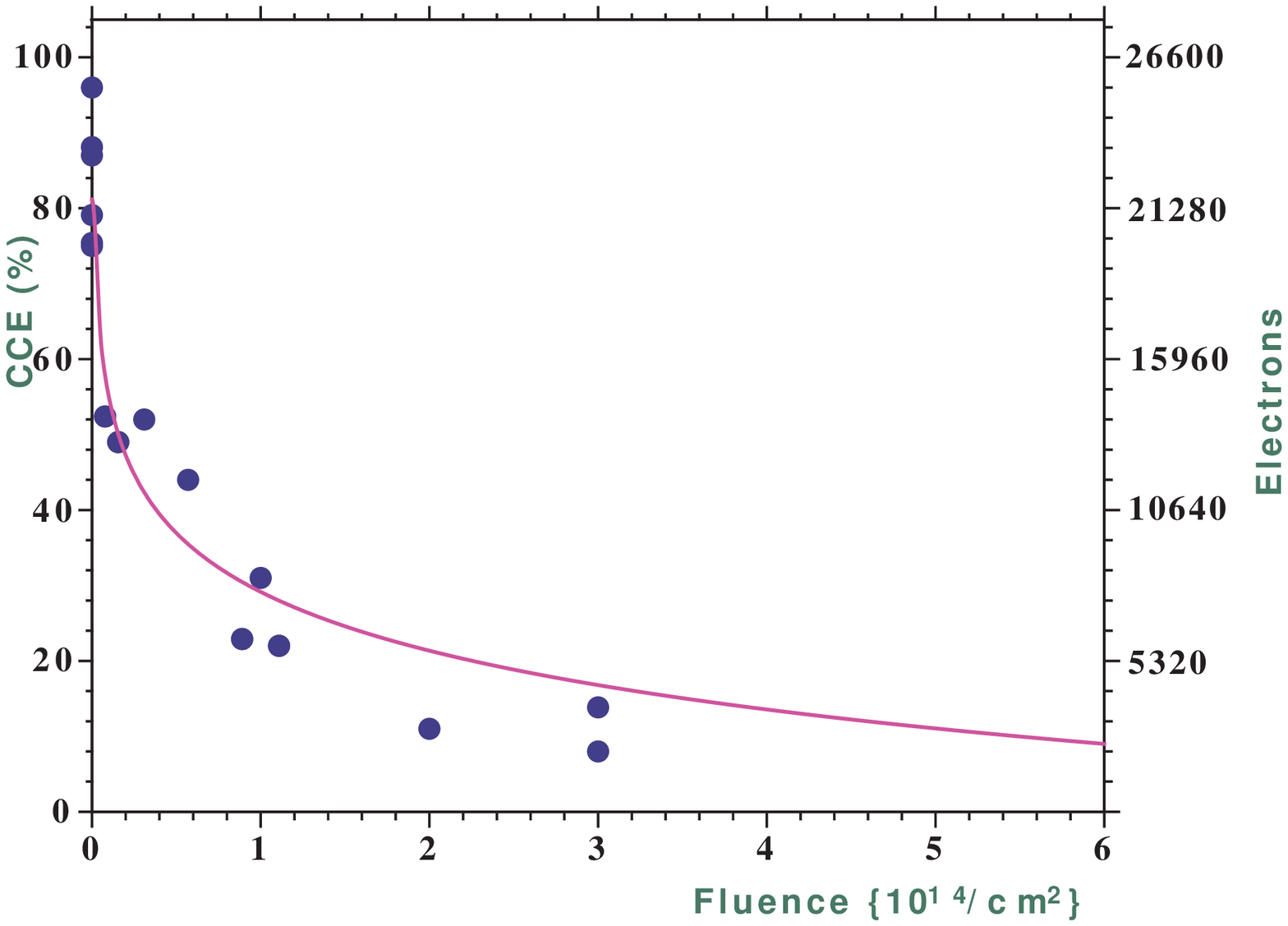}}}
\caption{The cce of 200$\mu$m thick GaAs detectors as a function of proton fluence, measured at 200V and 20$^{o}$C.
\label{fig:cceprotons}}
\end{figure}

\begin{figure}
\centering
\resizebox{.7\textwidth}{!}{\includegraphics{{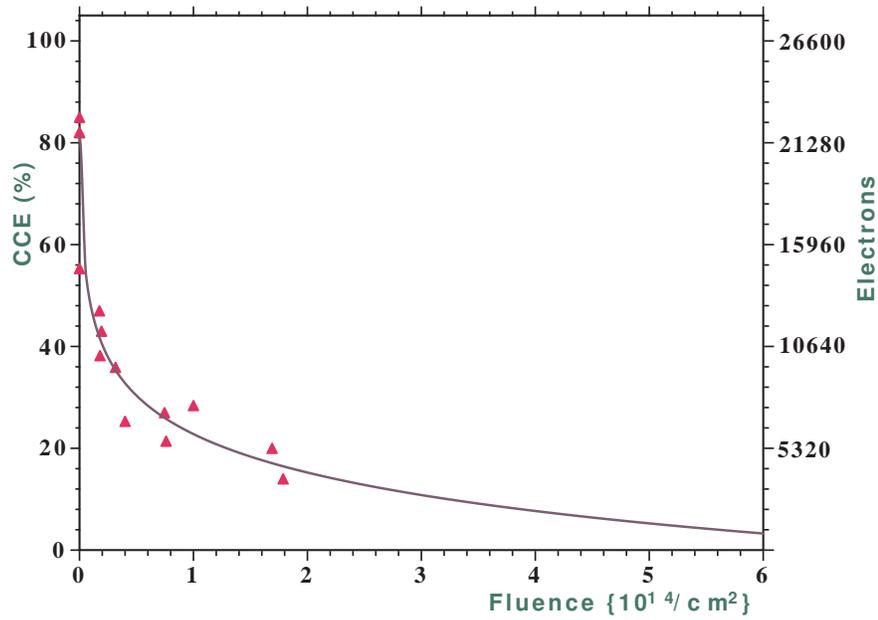}}}
\caption{The cce of 200$\mu$m thick GaAs detectors as a function of pion fluence, measured at 200V and 20$^{o}$C.
\label{fig:ccepions}}
\end{figure}

The charge collection efficiencies for 200$\mu$m thick detectors as a function of neutron, proton and pion fluence are shown in figures \ref{fig:cceneutrons}, \ref{fig:cceprotons} and \ref{fig:ccepions}. These were measured using minimum ionising particles (mips) from a $^{90}$Sr source at an applied bias of 200V. The reduction in cce with fluence occurs at a greater rate for pions and protons than for neutrons. A 10000 electron signal response to the $^{90}$Sr source, ($\approx$~40\% cce), is obtained after 1.4~$\times$~10$^{14}$~n/cm$^{2}$, 6.0~$\times$~10$^{13}$~p/cm$^{2}$, and only 3.0~$\times$~10$^{13}$~$\pi$/cm$^{2}$. The cce falls rapidly initially and then at a slower rate for higher fluences with the position of the change in gradient depending upon the particle type. 
In the ATLAS experiment at the LHC a particle flux of 0.12~$\times$~10$^{14}$~$\pi$/cm$^{2}$/year at the position of the GaAs wheels has been calculated \cite{gorfine} for full luminosity. This means that after only two years at full luminosity the pion fluence would be sufficient to reduce the signal from a 200$\mu$m thick detector to approximately 10000 electrons. The neutron flux is the only other significant contribution and this has been calculated to be a factor of 3 to 10 times less than the pions. The exact value depends upon the moderator used in the calorimeters. After the 10 years expected operational lifetime of ATLAS the neutron fluence will not be enough to reduce the signal to less than 10000 electrons.

\begin{figure}
\centering
\resizebox{.7\textwidth}{!}{\includegraphics{{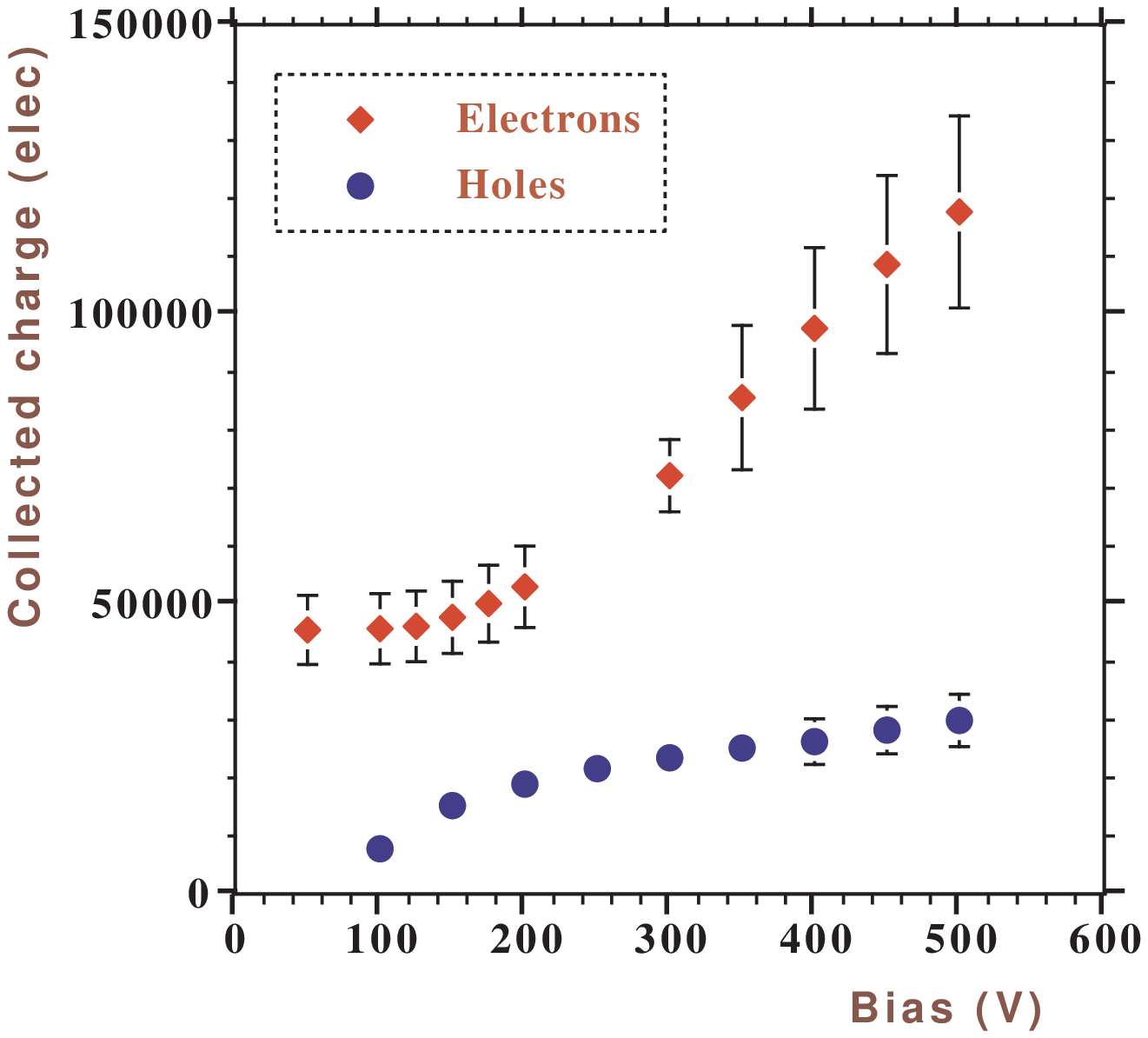}}}
\caption{The charge collected from electrons and holes as a function of bias for a 160$\mu$m thick GaAs detector after a fluence of 3~$\times$~10$^{14}$~p/cm$^{2}$.  
\label{fig:alphaehdata}}
\end{figure}

Irradiated 200$\mu$m thick detectors biased in excess of 200V demonstrate an increase in cce. For example for a 160$\mu$m thick detector irradiated to a fluence of 1.0~$\times$~10$^{14}$~p/cm$^{2}$ the signal increased from 8500 electrons at 200V to 12000 electrons at 400V. The reason for the increase in signal can be understood from the electron and hole collection data. Before irradiation, the signal due to the holes is greater than that due to the electrons, implying that the electrons are trapped more than the holes. The signal due to the holes however falls at a faster rate with fluence than that due to the electrons. This means that both electron and hole traps are introduced with irradiation, while the electron trapping is the less important after irradiation contrary to what is observed in unirradiated detectors. The electron signal, which is constant for a bias up to V$_{fd}$, rises at greater applied biases as shown in figure \ref{fig:alphaehdata}. The hole signal remains almost constant with increasing bias after its initial steep rise at V$_{fd}$. From this it is possible to conclude that field-enhanced electron detrapping may be present.

As the mip signal for a GaAs detector irradiated in excess of 1.0~$\times$~10$^{14}$~p/cm$^{2}$ appears to be independent of substrate thickness, even at an applied bias much larger than V$_{fd}$, it is proposed that the carriers have a mean free path less than or equal to the thickness of the thinnest detector tested, that is 115$\mu$m. The charge deposited by a mip, Q$_{dep}$ is given by 
\begin{equation}
Q_{dep} = \frac{dE}{dx} W
\end{equation}
where W is the width of the detector and dE/dx is the energy loss per unit length of detector traversed. If the carrier has a mean free path, $\lambda$, then the charge collected, Q$_{col}$, is
\begin{equation} 
Q_{col} = Q_{dep} \frac{\lambda}{W}
\end{equation}
thus 
\begin{equation} 
\lambda = \frac{Q_{col}}{dE/dx}
\end{equation}
The fitted values of $\lambda$, which is a function of the fluence, for proton and pion irradiation are given by equations (\ref{equ:lambdaprotons}) and (\ref{equ:lambdapions}), and show a similar fluence dependence. 
\begin{equation} 
\lambda = 96 \times \Phi^{-0.255}
\label{equ:lambdaprotons}
\end{equation}
\begin{equation} 
\lambda = 99 \times \Phi^{-0.252}
\label{equ:lambdapions}
\end{equation}

\section{Annealing studies}

The change in the reverse current of a detector irradiated to 1.0~$\times$~10$^{14}$~p/cm$^{2}$ has been measured over a period of time greater than 1 year. The sample was stored at room temperature, which varied throughout the year from 18$^{o}$C to 25$^{o}$C. All the current measurements were carried out at 20$^{o}$C with the same Keithley 237 voltage source measurement unit. The first current measurement was performed 20~days after the irradiation period due to the initial activity of the sample and holder preventing earlier handling. The value of the reverse current, I$_{r}$, at V$_{fd}$ reduced with time (t), in hours, with a logarithmic dependence given by
\begin{equation}
I_{r} \times 10^{7} = -0.063ln(t) + 2.4
\label{equ:itime}
\end{equation}
The value of V$_{fd}$ remained constant over this period. Some samples have shown an increase in cce over a similar period of annealing, however this represented at most an improvement of 8\% of the initial cce.

\begin{figure}
\centering
\resizebox{.7\textwidth}{!}{\includegraphics{{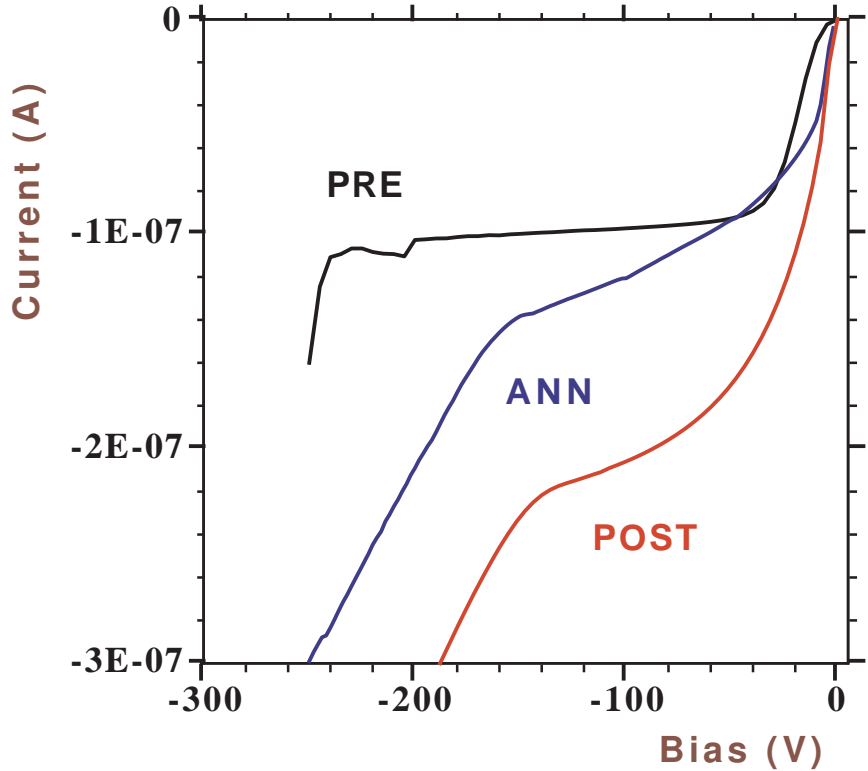}}}
\caption{The reverse current characteristics of a 180$\mu$m thick GaAs detector at 20$^{o}$C before and after 1.0~$\times$~10$^{14}$~p/cm$^{2}$ and after an anneal at 210$^{o}$C for 33 hours. 
\label{fig:210Cann}}
\end{figure}

\begin{figure}
\centering
\resizebox{.7\textwidth}{!}{\includegraphics{{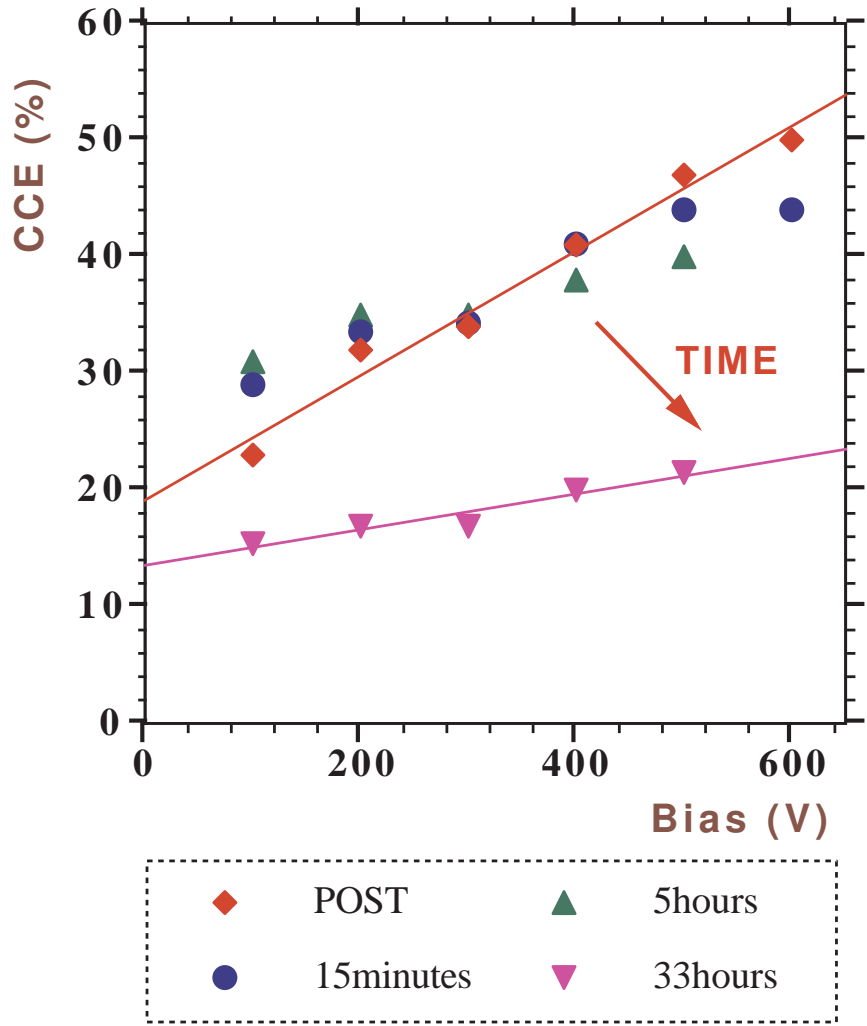}}}
\caption{The charge collection characteristics of a 180$\mu$m thick GaAs detector at 20$^{o}$C after 1.0~$\times$~10$^{14}$~p/cm$^{2}$ and after annealing at 210$^{o}$C for varying times.  
\label{fig:210cce}}
\end{figure}

Elevated temperature annealing, for periods ranging from 15~minutes to 33 hours, was performed on six irradiated detectors in an oven at 210$^{o}$C in air without bias. The reverse current and charge collection efficiency variations for a 180$\mu$m thick detector irradiated to 1.0~$\times$~10$^{14}$~p/cm$^{2}$ are shown in figures \ref{fig:210Cann} and \ref{fig:210cce}. The reverse current decreased after an 33 hour anneal from 0.22$\mu$A to 0.14$\mu$A (corresponding to 30nA/mm$^{2}$ to 20nA/mm$^{2}$) where the unirradiated value was 0.10$\mu$A (14nA/mm$^{2}$). The full depletion voltage remained unaffected by the annealing. For an applied bias of 100V the cce increased from 23\% to 30\% for annealing times up to 5 hours. As the bias increased, the increase in cce for the annealed detector was at a slower rate than in the pre-annealed case, resulting in a lower cce for the annealed detectors at biases in excess of 400V. For longer anneal times the cce decreased. For an anneal time of 33~hours, for example,  the cce was only 15\% at 100V rising to 22\% for a maximum applied bias of 500V.

The arsenic antisite defect in GaAs is known to anneal at temperatures in excess of 450$^{o}$C \cite{RTA}. As the irradiation could produce such antisite defects the effect of annealing at 450$^{o}$C was investigated. A flash-lamp anneal was used to illuminate the detectors for either 30,60,90, or 120 seconds. The effect was shown to be independent of the detector side which was illuminated, demonstrating that the whole bulk was being heated. The plateau reverse leakage current was reduced but so was the full depletion voltage. For an applied bias greater than V$_{fd}$ the annealed current was greater than before annealing with the gradient of this section of the current characteristic being quite large. The mip cce of the detector was increased for applied biases less than V$_{fd}$, however due to the large increase in leakage current above V$_{fd}$ it was not always possible to measure the charge collection at a higher applied bias. The increase in mip cce was due to an increase in electron mean free path with anneal time while the hole mean free path remained unchanged. A bias of 300V, which could be applied to a detector annealed at 450$^{o}$C for 120s after a fluence of 1.7~$\times$~10$^{14}$~$\pi$/cm$^{2}$, gave a mip cce of 71\%. The pre-irradiation cce was 95\% and the post-irradiation, pre-anneal value was 26\%.

\section{Indium doped samples}{\label{sec:indoped}}

Two detectors were fabricated with 500$\mu$m thick SI GaAs doped with indium \cite{mcp}. These were tested before and after an exposure of 1.4~$\times$~10$^{14}$ 24GeV/c protons/cm$^{2}$.
The reverse current of the detectors before irradiation was 30nA/mm$^{2}$, less than for the 500$\mu$m thick SI-U GaAs detectors. After the irradiation the value of V$_{fd}$ was reduced to only 220V, compared to 340V for the undoped sample. The current at the applied bias of V$_{fd}$ was 0.4$\mu$A (57nA/mm$^{2}$) which was higher than that of the undoped sample which was 0.22$\mu$A.
The charge collection efficiency for the indium doped samples was less that that of the undoped samples before irradiation by 20\%. After the irradiation the signal fell to 10\% (6650 electrons) compared to 17\% (11305 electrons), respectively, for an applied bias of 1000V.

\section{Conclusions}

The changes in reverse current and charge collection efficiency for SI-U GaAs detectors have been studied after irradiations with 1MeV neutrons, 24GeV/c protons and 300MeV/c pions. While the leakage current variations are slight and pose little problem to the operation of the detectors, the fall in cce gives major cause for concern. At a given high fluence, the reduction is greater for charged hadrons than for neutrons (this difference is discussed in detail in reference \cite{terry}). The cce falls rapidly in the first 3~$\times$~10$^{13}$~p($\pi$)/cm$^{2}$. The reduction has been shown to be due to both electrons and holes being trapped, with the hole mean free path affected much more dramatically. For an applied bias greater than V$_{fd}$ the cce increases due to the increased signal from the electrons. 
Room temperature annealing has been identified in the reverse current characteristics and a slight improvement has also been seen in the cce.
Annealing at elevated temperatures has shown beneficial effects in both leakage current and cce. 
The effect of proton irradiation on indium-doped GaAs detectors has shown the material to be no more radiation-hard than SI-U GaAs. Further tests on different samples should be performed.

\section{Acknowledgements}

The authors would like to thank F.McDevitt, A.Meikle and F.Doherty for technical support and all those at the ISIS, CERN PS and PSI facilities during the irradiation runs. One of us (R.Bates) gratefully acknowledges the support received through a CASE postgraduate studentship from RAL. The results obtained within the RD8 collaboration are from work partly funded by PPARC (UK), INFN (Italy), and the BMFD (Germany).

\end{document}